%% file: main.tex
\documentclass[sigconf]{acmart}

\usepackage{booktabs}  
\usepackage{array}    
\usepackage{tabularx}  
\usepackage{graphicx}   
\usepackage{makecell}   
\usepackage{multirow}
\usepackage{caption}
\usepackage{url}
\usepackage{hyperref}
\usepackage{subcaption}
\usepackage{listings}
\usepackage{enumitem}
\usepackage{makecell}

\usepackage{stfloats}      

\usepackage{colortbl}
\usepackage{multirow}

\AtBeginDocument{%
  }

\copyrightyear{2026}
\acmYear{2026}
\setcopyright{cc}
\setcctype{by-nc-nd}

\acmConference[ICAIF '26]{7th ACM International Conference on AI in Finance}{November 14--17, 2026}{Milan, Italy}
\acmBooktitle{7th ACM International Conference on AI in Finance (ICAIF '26), November 14--17, 2026, Milan, Italy}
\acmDOI{10.1145/3858063.3858815}
\acmISBN{979-8-4007-3087-0/2026/11}

\begin{document}

\title{On-Chain Behavioral Signatures for Classifying Centralized Exchange Addresses on Ethereum Using Proof-of-Reserves}

\author{Seongchan Park}
\orcid{0009-0007-8575-722X}
\affiliation{%
  \institution{Dunamu}
  \city{Seoul}
  \country{Republic of Korea}
}
\email{chance@dunamu.com}

\author{Hyunho Jeon}
\orcid{0009-0001-5153-2520}
\affiliation{%
  \institution{Dunamu}
  \city{Seoul}
  \country{Republic of Korea}
}
\email{jayden.jeon@dunamu.com}

\author{Byungsoo Kang}
\orcid{0009-0009-2259-3226}
\affiliation{%
  \institution{Dunamu}
  \city{Seoul}
  \country{Republic of Korea}
}
\email{bubba@dunamu.com}

\author{Gaeun Kim}
\orcid{0009-0001-1035-3788}
\affiliation{%
  \institution{Dunamu}
  \city{Seoul}
  \country{Republic of Korea}
}
\email{ganni@dunamu.com}

\author{Minseob So}
\orcid{0009-0004-2355-2681}
\affiliation{%
  \institution{Dunamu}
  \city{Seoul}
  \country{Republic of Korea}
}
\email{seop@dunamu.com}

\author{Seongyong Kim}
\orcid{0009-0007-0893-6179}
\affiliation{%
  \institution{Dunamu}
  \city{Seoul}
  \country{Republic of Korea}
}
\email{levi@dunamu.com}

\renewcommand{\shortauthors}{Park et al.}

\input{0_abstract}




\maketitle

\vspace{-0.8mm}
\input{1_introduction}
\input{2_relatedwork}
\input{3_dataset}
\input{4_method}
\input{5_result}
\input{6_discussion}
\input{7_conclusion}

\begin{acks}
The authors thank the anonymous reviewers for their constructive feedback, as well as Daehyun Kim, the \href{https://datalab.upbit.com/}{Upbit DataLab} team, and \href{https://www.lambda256.io/}{Lambda256} for their support during this research.
\end{acks}

\bibliographystyle{ACM-Reference-Format}
\bibliography{references}

\end{document}

%% file: 0_abstract.tex
\begin{abstract}

Centralized exchanges (CEXs) are key gateways between traditional finance and digital asset markets, yet identifying their blockchain addresses remains challenging. Existing approaches often rely on opaque attribution labels or treat exchanges as secondary classification targets, leaving generalization to unseen exchanges insufficiently examined. This study develops a classification framework using Proof-of-Reserves (PoR) disclosures, with 3,033,857 Ethereum addresses from 15 CEXs and five Non-CEX categories. We derive 58 on-chain behavioral features and evaluate LightGBM against logistic regression (LR) under a leave-one-exchange-out protocol. Across five lookback windows (30–730 days), LightGBM consistently outperformed LR, with macro PR-AUC increasing from 0.6611 to 0.9594 using the selected 17 features, compared with 0.7697 for LR at 730 days. Robustness checks showed comparable performance when feature selection and hyperparameter optimization were repeated across splits and held-out exchanges, while exchange-balanced reweighting improved macro PR-AUC. SHAP analysis and feature-group ablation identified gas behavior, ERC-20 token activity, and internal transactions as key predictive signals in the constructed dataset. These findings support interpretable CEX address screening for blockchain monitoring, while the released feature dataset enables reproducible follow-up research.\footnote{\url{https://github.com/dunamu-labs/cex-address-classification}}

\end{abstract}

%% file: 1_introduction.tex
\section{Introduction}

Centralized exchanges (CEXs) occupy a unique structural position in the cryptocurrency ecosystem. Beyond facilitating trade execution, they mediate fiat–crypto conversion, support stablecoin flows, and often function as exit points for illicit funds~\cite{makarov2020trading,reuter2025decrypting,song2024identifying}. Accordingly, characterizing CEX-controlled addresses is a fundamental prerequisite for market monitoring, risk assessment, and anti-money laundering (AML) investigations~\cite{hoang2022loaded,wu2023toward, iftikhar2025systemic}. Despite this importance, identifying these entities remains challenging because the pseudonymity of blockchain addresses obscures both wallet ownership and operational roles from the ledger alone. Prior research has approached this challenge using structural methods such as address clustering and entity classification~\cite{victor2020address, garin2023machine, meiklejohn2013fistful}, and through behavioral signatures for detecting malicious actors~\cite{lorenz2020machine,song2024identifying}, including scams and phishing schemes~\cite{vasek2015there,chen2021phishing}. However, CEX addresses are still frequently treated as a secondary licit category within broader frameworks rather than as a primary classification target~\cite{weber2019anti}, leaving their specific behavioral patterns underexplored. Moreover, labels from blockchain explorers and intelligence platforms rely on proprietary attribution logic, which may introduce label noise and limit the reproducibility of supervised learning studies.

\begin{figure*}[th]
    \centering
    \includegraphics[width=1.00\linewidth]{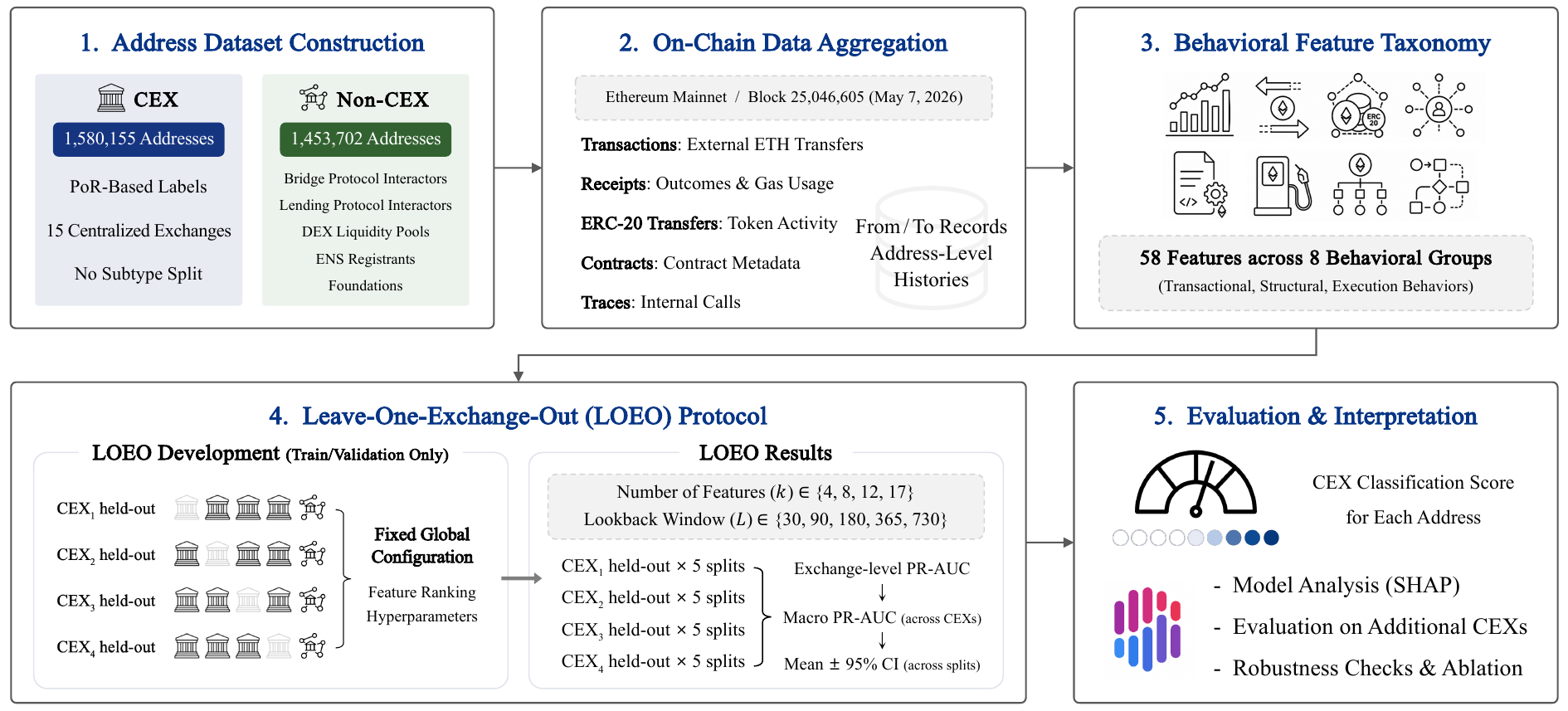}
    \caption{Overview of the proposed CEX address classification framework with leave-one-exchange-out (LOEO) evaluation.}
    \label{fig:process_overview}
\end{figure*}

To address these limitations, we propose a research framework, illustrated in Figure~\ref{fig:process_overview}, that leverages high-confidence labels derived from Proof-of-Reserves (PoR) disclosures. We construct a dataset consisting of 3,033,857 unique wallet addresses, incorporating positive labels from 15 centralized exchanges and a heterogeneous Non-CEX set spanning bridge interactors, lending interactors, DEX liquidity pools, Ethereum Name Service (ENS) registrants, and foundations. Our pipeline aggregates on-chain records from the Ethereum mainnet up to block 25,046,605 (May 7, 2026) and transforms them into a structured behavioral taxonomy of 58 domain-specific features, organized into eight groups capturing transactional, structural, and execution patterns. To examine how observation depth affects classification performance, we evaluate the pipeline across lookback windows from 30 to 730 days. A validation-only selection procedure reduces the 58 features to 17, which are evaluated with LightGBM and logistic regression (LR) under a leave-one-exchange-out (LOEO) protocol~\cite{ke2017lightgbm}, followed by SHAP-based interpretation and feature-group ablation analysis~\cite{lundberg2017unified}.

Our results show that CEX identification benefits from longer lookback windows and broader feature coverage: macro PR-AUC increased from 0.6003 with four features and a 30-day lookback to 0.9594 with 17 features and a 730-day lookback, outperforming LR (0.7697) under the same final configuration. Beyond these results, the study makes three contributions. First, it constructs an address-level dataset combining PoR-based CEX labels with heterogeneous Non-CEX groups, and releases a category-labeled feature dataset to support follow-up research. Second, it evaluates cross-exchange generalization using a LOEO protocol that prevents exchange overlap between model development and held-out CEX test samples. Third, it examines model behavior through SHAP analysis and robustness checks, followed by feature-group ablation to assess the predictive contributions of on-chain features.


%% file: 2_relatedwork.tex
\section{Related Work}

\subsection{Blockchain Address Classification}

The identification of real-world entities behind pseudonymous blockchain addresses has evolved from static transaction linking to dynamic behavioral analysis. Early work on Bitcoin introduced heuristic-based clustering to group addresses by shared spending authority~\cite{meiklejohn2013fistful}. Subsequent studies demonstrated that structural properties can help distinguish exchanges from other entities~\cite{ron2013quantitative}. On Ethereum, these ideas were adapted to the account-based model through patterns such as deposit address reuse~\cite{victor2020address}. This line of research was further enriched by complex network modeling, which incorporates temporal and multiplicity characteristics to capture account-level fund flows~\cite{lin2020modeling}. More broadly, these developments reflect a paradigm shift from manual heuristics toward learned behavioral signatures derived from transaction histories~\cite{garin2023machine}.

In parallel, AML research has examined the detection of illicit activities such as scams, phishing, and Ponzi schemes~\cite{vasek2015there,chen2021phishing,chen2018detecting}. These efforts have progressed from node-level classification, where exchanges are treated as one of several licit categories~\cite{weber2019anti,lorenz2020machine}, to subgraph-level detection of laundering patterns such as peeling chains~\cite{song2024identifying}. A recurring finding is that exchanges are not merely passive service nodes, but major intermediaries and eventual exit points in laundering processes, making them structurally important for understanding and tracing illicit fund flows~\cite{song2024identifying,wu2023toward}. 

Despite this central role, CEX identification has rarely been studied as a primary classification target on Ethereum. Existing approaches typically treat exchanges as a general licit category within broader AML frameworks or infer them through heuristic attribution, rather than modeling them as a distinct address class with dedicated behavioral features. Our work addresses this gap by isolating the behavioral signatures of centralized exchanges using high-confidence labels derived from proof-of-reserves disclosures. 

\subsection{On-Chain Behavioral Patterns of Exchanges}
\label{sec:behavioral_patterns}

The operational logic of centralized platforms produces distinctive on-chain footprints through the systematic consolidation of assets from customer-specific deposit addresses into exchange-controlled wallets. This process creates intermediary address flows that can be identified by analyzing transfer value distributions and the time intervals between receiving and forwarding~\cite{victor2020address}. At the network level, ERC-20 token transfer graphs exhibit star or hub-and-spoke structures where exchange addresses emerge as central hubs with high in-degree~\cite{victor2019measuring}. Similarly, PageRank and degree centrality analyses show that exchange-related nodes occupy highly central positions within Ethereum money flow graphs~\cite{chen2020understanding}. These addresses have also been identified as structurally important nodes in Ethereum token networks. Together with smart contract addresses, they are associated with rapid fragmentation under node-removal experiments, suggesting that overall network connectivity depends on a small number of centralized intermediaries~\cite{de2022structural}. 

Beyond structural properties, prior work shows that exchange activity can be characterized through broader behavioral signals. On-chain profiling based on temporal activity, gas price usage, and transaction graph features provides meaningful signals for distinguishing different types of entities~\cite{beres2021blockchain}. Furthermore, centralized exchanges play a key role in linking fiat and crypto ecosystems through stablecoin flows. In particular, Binance and Coinbase have been shown to differ in their respective specialization in stablecoin on-ramping and off-ramping~\cite{reuter2025decrypting}. 

Taken together, these findings demonstrate that exchanges are identifiable not only through local transactional heuristics, such as deposit-and-forward behavior, but also through broader temporal, fee-related, and structural patterns. This motivates our feature design in Section~\ref{sec:feature_design}, where we organize these on-chain signals into a unified behavioral taxonomy for CEX classification.

%% file: 3_dataset.tex
\section{Wallet Dataset}

\subsection{CEX Wallets}

A key prerequisite for CEX wallet classification is a set of high-confidence positive labels. While third-party blockchain explorers and intelligence platforms (e.g., Etherscan, Arkham Intelligence) annotate exchange-related accounts, the underlying attribution criteria are not always publicly documented. This limited transparency can make it difficult to independently verify labels or reproduce the labeling process for supervised learning.

To address this issue, we use PoR disclosures published directly by exchanges as the primary source of CEX labels. Following a series of high-profile crypto firm failures in 2022, many centralized exchanges adopted PoR to improve transparency around reserve holdings~\cite{vidal2025centralized}. These materials may include reserve snapshots, publicly attributable accounts, and signed messages that support cryptographic verification of address ownership. For example, Binance and OKX release account lists together with reserve information or signed ownership proofs, while Bybit and KuCoin offer public reserve pages supported by audit documentation.

For this study, we collected addresses from 15 centralized exchanges as of May 2026. Although certain platforms provide granular data on operational subtypes (e.g., hot, cold, and deposit wallets), such information is inconsistent across the board. To maintain consistency, we do not subdivide these accounts, assigning instead a single positive label to all collected CEX accounts.

\subsection{Non-CEX Wallets}

To construct a diverse negative dataset, we included five Non-CEX categories: bridge protocol interactors, lending protocol interactors, DEX liquidity pools, ENS registrants, and foundation addresses. Their heterogeneous on-chain behaviors allow the classifier to distinguish CEX addresses from a broad range of Non-CEX groups.

\vspace{2mm}
\noindent \textbf{Bridge Protocol Interactors:}
We identified addresses initiating transactions with L1 native bridge contracts across 43 L2 networks, including Arbitrum, Base, Optimism, and Polygon. Among 3,628,241 candidate addresses, only 23 (0.0006\%) overlapped with known CEX PoR wallets and were excluded. To reduce incidental usage, we retained addresses that interacted with bridges associated with two or more distinct L2 networks.

\vspace{2mm}
\noindent \textbf{Lending Protocol Interactors:}
We extracted addresses interacting with contracts across four major Ethereum lending protocols (Aave V2, Aave V3, Compound V2, and MakerDAO). Out of 497,417 candidate addresses, only 17 (0.0034\%) coincided with CEX PoR wallets; these overlaps were excluded from the Non-CEX set. This category incorporates on-chain borrowing and collateral management activity into our negative class.

\vspace{2mm}
\noindent \textbf{DEX Liquidity Pools:}
We targeted liquidity pool addresses from major protocols, including Uniswap, SushiSwap, and PancakeSwap. By tracing events such as \texttt{PoolCreated} or \texttt{PairCreated} emitted by factory contracts, we extracted the corresponding pool addresses. This approach incorporates characteristic activity of automated market maker (AMM)-based liquidity pools into our negative class.

\vspace{2mm}
\noindent \textbf{ENS Registrants:}
We used ENS registration data as a proxy for individual or non-institutional participants. Specifically, we extracted the \texttt{owner} field from \texttt{NameRegistered} events emitted by both legacy and current \texttt{ETHRegistrarController} contracts. To reduce potential label contamination, we lowercased ENS names and excluded entries containing exchange-related keywords (e.g., binance, coinbase, and okx). Although this heuristic cannot rule out all institutional or exchange-affiliated actors, it provides a scalable means of identifying likely non-institutional ENS registrants.

\vspace{2mm}
\noindent \textbf{Foundations:}
We collected addresses associated with 139 distinct token foundations from their official project documentation. These addresses capture organizationally managed on-chain activity outside centralized exchanges and serve as a complementary negative category alongside decentralized protocol-related addresses.

\begin{table}[!b]
\centering
\caption{Distribution of the labeled dataset, comprising CEX and Non-CEX addresses on the Ethereum mainnet.}
\label{tab:entity_distribution}

\begin{tabularx}{\linewidth}{
    >{\raggedright\arraybackslash}X
    >{\raggedleft\arraybackslash}r
    >{\raggedleft\arraybackslash}r
}
\toprule
\textbf{Class / Category}
& \textbf{Addresses}
& \textbf{Share (\%)} \\
\midrule

\textbf{CEX}
& \textbf{1,580,155}
& \textbf{52.08} \\

\quad Binance
& 1,468,505
& 48.40 \\

\quad Bitget
& 60,788
& 2.00 \\

\quad OKX
& 48,220
& 1.59 \\

\quad Korbit
& 2,433
& 0.08 \\

\quad Others$^\dagger$
& 209
& 0.01 \\

\midrule

\textbf{Non-CEX}
& \textbf{1,453,702}
& \textbf{47.92} \\

\quad Bridge Protocol Interactors
& 628,771
& 20.73 \\

\quad Lending Protocol Interactors
& 457,387
& 15.08 \\

\quad DEX Liquidity Pools
& 277,597
& 9.15 \\

\quad ENS Registrants
& 87,690
& 2.89 \\

\quad Foundations
& 2,257
& 0.07 \\

\midrule

\textbf{Total Unique Addresses}
& \textbf{3,033,857}
& \textbf{100.00} \\

\bottomrule
\addlinespace

\multicolumn{3}{r}{
    \parbox{0.9\linewidth}{
        \raggedleft
        $^\dagger$ \footnotesize
        Includes Bybit, CoinDCX, Huobi, MEXC, KuCoin, Gemini,
        BingX, INDODAX, Phemex, Qmall, and Bitfinex.
    }
}
\end{tabularx}
\end{table}

\subsection{Final Address-Level Dataset}

After collecting the CEX and Non-CEX address sets, valid Ethereum addresses in hexadecimal format beginning with \texttt{0x} were retained. Each address was required to have at least one observed transaction on the Ethereum mainnet before the global cutoff, establishing an observed transaction history for feature extraction. While cross-referencing confirmed no overlap between the CEX and Non-CEX classes, addresses appearing in multiple Non-CEX categories were deduplicated. As a result, each address was assigned to a single category, and the two classes remained mutually exclusive. The resulting dataset contains 3,033,857 unique addresses, comprising 1,580,155 CEX (52.08\%) and 1,453,702 Non-CEX (47.92\%) addresses. Although the two classes are similar in size, the CEX class is concentrated in Binance, largely reflecting differences in the composition and scale of publicly available PoR disclosures across exchanges. Eleven exchanges with relatively few disclosed addresses were grouped under \textit{Others} in Table~\ref{tab:entity_distribution}.

%% file: 4_method.tex
\section{On-Chain Behavioral Features}

\begin{table*}[th]
\centering
\caption{Catalog of 58 designed features across eight behavioral groups, with the 17 selected features shown in bold.}
\label{tab:full_features_final}

\footnotesize
\renewcommand{\arraystretch}{1.25}
\setlength{\tabcolsep}{2pt}

\begin{tabularx}{0.95\textwidth}{
    >{\hspace{5pt}
      \hsize=1.05\hsize
      \linewidth=\hsize
      \raggedright\arraybackslash}X
    >{\hsize=0.95\hsize
      \linewidth=\hsize
      \raggedright\arraybackslash}X
    !{\color{gray!40}\vrule width 0.8pt}
    >{\hspace{5pt}
      \hsize=1.05\hsize
      \linewidth=\hsize
      \raggedright\arraybackslash}X
    >{\hsize=0.95\hsize
      \linewidth=\hsize
      \raggedright\arraybackslash}X
}
\toprule
\textbf{Feature Name} & \textbf{Description} & \textbf{Feature Name} & \textbf{Description} \\
\midrule

\multicolumn{2}{>{\columncolor{gray!20}}c}{\textbf{Group 1: Transaction Statistics (14)}} & \multicolumn{2}{>{\columncolor{gray!20}}c}{\textbf{Group 5: Contract Interactions (6)}} \\
\texttt{tx\_count\_\{in,out\}} & In- \& out-tx counts & \texttt{is\_contract} & Smart contract account flag \\
\texttt{tx\_count\_total} & Total tx count & \textbf{\texttt{pct\_tx\_to\_contract}} & \% of out-tx to contracts \\ 
\texttt{eth\_\{sent,recv\}\_total} & Total ETH sent \& received & \texttt{unique\_contracts\_called} & Unique contracts called \\
\textbf{\texttt{eth\_sent\_median}} & Median outbound ETH & \texttt{has\_uniswap\_interaction} & Uniswap interaction flag \\
\textbf{\texttt{eth\_sent\_p95}} & 95th percentile of outbound ETH & \texttt{uniswap\_tx\_count} & Uniswap tx count \\
\texttt{eth\_sent\_std} & Std.\ dev.\ of outbound ETH & \texttt{uniswap\_tx\_ratio} & Uniswap tx / total out-tx ratio \\
\textbf{\texttt{eth\_recv\_max}} & Maximum single ETH receipt & \multicolumn{2}{>{\columncolor{gray!20}}c}{\textbf{Group 6: Gas Usage \& Fees (5)}} \\
\texttt{active\_days} & Days with $\geq$ 1 tx & \textbf{\texttt{avg\_priority\_fee\_gwei}} & Avg.\ priority fee (Gwei) \\
\texttt{wallet\_age\_days} & Days from 1st tx to cutoff & \textbf{\texttt{avg\_gas\_used}} & Avg.\ gas used \\
\texttt{tx\_per\_active\_day} & Avg.\ tx per active day & \texttt{eip1559\_tx\_ratio} & EIP-1559 tx ratio \\
\texttt{failed\_tx\_count} & Reverted tx count & \textbf{\texttt{avg\_fee\_eth}} & Avg.\ tx fee (ETH) \\
\texttt{failed\_tx\_ratio} & Reverted tx ratio & \texttt{total\_fees\_eth} & Total gas fees (ETH) \\
\multicolumn{2}{>{\columncolor{gray!20}}c}{\textbf{Group 2: ETH Flow Patterns (5)}} & \multicolumn{2}{>{\columncolor{gray!20}}c}{\textbf{Group 7: Internal Transactions (7)}} \\
\texttt{eth\_inflow\_outflow\_ratio} & Received / sent ETH ratio & \texttt{\textbf{internal\_tx\_count\_}\{in,\textbf{out}\}} & Internal in- \& out-tx counts \\
\textbf{\texttt{eth\_net\_flow}} & Net ETH flow (received $-$ sent) & \texttt{internal\_tx\_count\_total} & Total internal tx count \\
\texttt{large\_tx\_fraction\_\{in,out\}} & Fraction of in- \& out-tx $>$ 10 ETH & \texttt{internal\_eth\_sent} & Total ETH via internal calls \\
\textbf{\texttt{zero\_value\_tx\_ratio}} & Zero-value tx ratio & \texttt{internal\_to\_external\_ratio} & Internal out-tx / out-tx ratio \\
\multicolumn{2}{>{\columncolor{gray!20}}c}{\textbf{Group 3: ERC-20 Token Activity (6)}} & \textbf{\texttt{avg\_subtraces}} & Avg.\ subtraces per internal out-tx \\
\texttt{\textbf{erc20\_unique\_tokens\_}\{\textbf{sent},recv\}} & Unique tokens sent \& received & \texttt{error\_rate\_internal} & Error rate of internal calls \\
\texttt{\textbf{erc20\_tx\_count\_}\{\textbf{in},out\}} & ERC-20 in- \& out-tx counts & \multicolumn{2}{>{\columncolor{gray!20}}c}{\textbf{Group 8: Composite Patterns (7)}} \\
\texttt{erc20\_tx\_count\_total} & Total ERC-20 tx count & \texttt{first\_outbound\_delay\_h} & First-in to first-out delay (hours) \\
\textbf{\texttt{erc20\_to\_eth\_tx\_ratio}} & ERC-20 tx / total tx ratio & \textbf{\texttt{forwarding\_value\_ratio}} & Sent / received ETH ratio \\
\multicolumn{2}{>{\columncolor{gray!20}}c}{\textbf{Group 4: Counterparty Patterns (8)}} & \textbf{\texttt{stablecoin\_tx\_ratio}$^\dagger$} & Stablecoin ERC-20 transfer share \\
\texttt{\textbf{unique\_}\{sender,\textbf{receiver}\}} & Unique sender \& receiver addresses & \texttt{token\_diversity\_entropy} & Shannon entropy of token distribution \\
\texttt{hhi\_\{sender,receiver\}} & HHI of in- \& out-tx counts & \texttt{batch\_tx\_ratio} & Fraction of blocks with $\geq$ 3 out-tx \\
\texttt{gini\_\{sender,receiver\}} & Gini of in- \& out-tx counts & \texttt{same\_block\_cpty\_max} & Maximum same-block counterparties \\
\texttt{top1\_\{sender,receiver\}\_share} & Top sender \& receiver shares & \texttt{activity\_lifespan\_ratio} & Active period / wallet age ratio \\
\bottomrule
\addlinespace
\multicolumn{4}{r}{$^\dagger$ \footnotesize Stablecoins included: USDT, USDC, DAI, USDe, and USD1 (top five by Ethereum ERC-20 transfer volume).}
\end{tabularx}
\end{table*}

\subsection{Data Source and Lookback Construction}
Comprehensive activity records were obtained from the Ethereum mainnet through Nodit by Lambda256, up to a fixed cutoff at block 25,046,605. The raw dataset was organized into five primary tables: Transactions, Receipts, ERC-20 Transfers, Contracts, and Traces. \texttt{Transactions} and \texttt{Receipts} supported the analysis of external ETH transfers, execution outcomes, and gas usage. These records were complemented by \texttt{ERC-20 Transfers} for tracking token movements and \texttt{Contracts} metadata for identifying smart contract accounts and interaction targets. In addition, \texttt{Traces} provided information on internal calls and execution complexity at the virtual machine level. 
For each labeled address, relevant records were extracted from the five tables to capture both incoming and outgoing activity, whether the address appeared as a sender or a receiver. Using May 7, 2026 as the global reference date, we identified the most recent transaction observed for each address. To enable efficient batch aggregation, we aligned each address's lookback anchor to the first day of the month following its last transaction. The five lookback windows (30, 90, 180, 365, and 730 days) were defined relative to this aligned date, with on-chain features computed separately for each address and window. Consequently, windows of the same length may cover different calendar periods across addresses.



\subsection{Feature Taxonomy}
\label{sec:feature_design}

To represent address-level on-chain behavior, we designed 58 features organized into eight behavioral groups, as shown in Table~\ref{tab:full_features_final}. The taxonomy covers transaction statistics, ETH flow patterns, ERC-20 token activity, counterparty structure, contract interactions, gas behavior, internal transactions, and composite behavioral patterns.

\textbf{\textit{Groups 1--2}} quantify transaction volume, activity density, and ETH flow asymmetry. They are motivated by deposit-and-forward heuristics, in which discrepancies between inflow and outflow volumes and forwarding delays can indicate intermediary activity~\cite{chen2021phishing, victor2020address,garin2023machine}. \textbf{\textit{Groups 3--4}} characterize ERC-20 token activity and counterparty concentration, reflecting the hub-and-spoke topology associated with exchange-controlled nodes~\cite{victor2019measuring,chen2020understanding,de2022structural}. Concentration and inequality metrics, such as the Herfindahl--Hirschman Index (HHI) and Gini coefficients, distinguish broad interaction patterns from concentrated counterparty structures~\cite{chemaya2025dataset}. \textbf{\textit{Groups 5--7}} capture execution-level behavior, including smart-contract usage, gas pricing, and internal transaction patterns, following prior studies of systematic fee usage and multi-step transaction structures~\cite{beres2021blockchain}. Finally, \textbf{\textit{Group 8}} combines multiple primitives into higher-order indicators of forwarding behavior, stablecoin usage, token diversity, and activity lifespan. These features are partly motivated by evidence that exchanges serve as important conduits for stablecoin-based on- and off-ramping between fiat and crypto ecosystems~\cite{reuter2025decrypting}.

\begin{figure*}[th]
    \centering
    \includegraphics[width=1.00\linewidth]{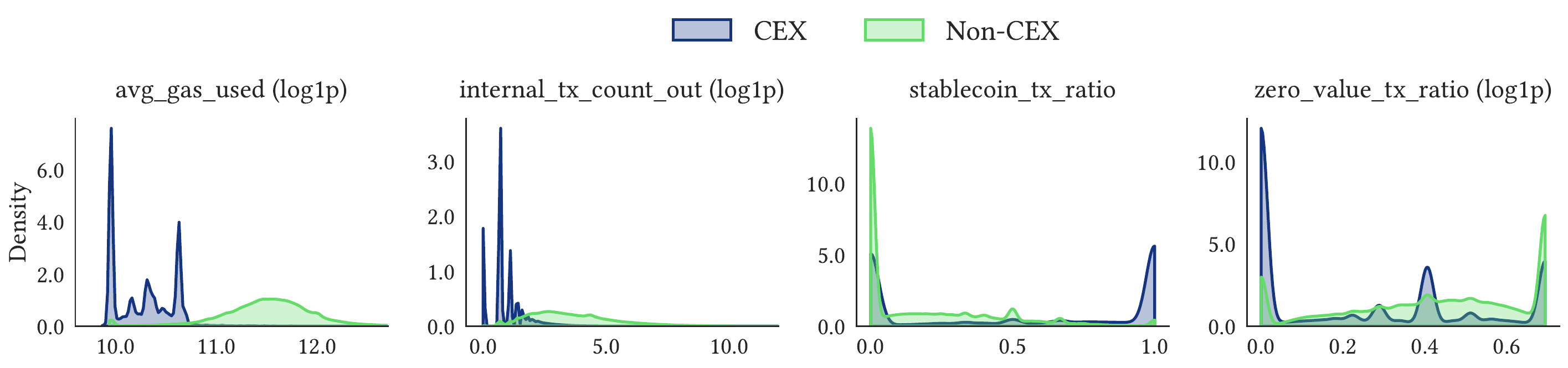}
    \caption{Kernel density estimates of four representative features for CEX and Non-CEX addresses (730-day lookback window).}
    \label{fig:core_feature_stats}
\end{figure*}

\subsection{Descriptive Analysis}
The empirical patterns of selected features are examined over the 730-day horizon, which provides broader behavioral coverage than shorter periods for less active addresses. Figure~\ref{fig:core_feature_stats} presents kernel density estimates for four representative features, with logarithmic transformations applied where indicated. Both \texttt{avg\_gas\_used} and \texttt{internal\_tx\_count\_out} show CEX addresses concentrated at lower values, whereas the Non-CEX group displays broader, right-shifted distributions. For \texttt{stablecoin\_tx\_ratio}, Non-CEX values are clustered near zero, while the CEX group shows substantial density near both zero and one. Lastly, \texttt{zero\_value\_tx\_ratio} exhibits distinct multi-peaked patterns across the two classes. These differences suggest that gas usage, internal call activity, and transaction composition provide complementary signals, while the observed overlap highlights the need for multivariate evaluation.

%% file: 5_result.tex
\section{Method}

\subsection{Leave-One-Exchange-Out Protocol}
\label{sec:loeo_protocol}


A conventional random split may place addresses from the same exchange in both training and test sets, allowing models to exploit exchange-specific patterns rather than learn transferable behavioral signatures. We therefore adopt a leave-one-exchange-out (LOEO) protocol that keeps each held-out CEX disjoint from model development, providing a stricter test of generalization to unseen exchanges. Binance, Bitget, OKX, and Korbit were sufficiently represented for LOEO evaluation, while smaller exchanges grouped under \textit{Others} were evaluated separately. In each LOEO setting, one of the four exchanges was held out for testing, while the other three provided positive samples for training and validation.


Data partitioning was performed at the unique-address level rather than the observation level, ensuring consistent assignment across all five lookback windows (30, 90, 180, 365, and 730 days). We first created five stratified 60:20:20 train--validation--test splits across all eligible addresses, preserving the proportions of individual exchanges and Non-CEX source categories. For each LOEO task, the training and validation sets consisted of the corresponding folds from the three non-held-out exchanges and the Non-CEX pool. Evaluation included all addresses from the held-out exchange and the Non-CEX test fold, excluding test-fold addresses from the remaining exchanges to ensure exchange-disjointness. Combining four held-out exchanges, five splits, and five lookback windows produced $4 \times 5 \times 5 = 100$ base LOEO configurations. For each held-out exchange and lookback window, the area under the precision--recall curve (PR-AUC) was computed for each split and averaged across the five splits.

\subsection{Feature Selection and Model Training}

We employed LightGBM as our primary classifier for its strong performance on tabular data, native handling of missing values, and efficient SHAP-based interpretation~\cite{ke2017lightgbm, grinsztajn2022tree, lundberg2017unified}. Missing values in count, sum, and binary features were replaced with zero to represent inactivity, whereas those in ratio, average, and concentration features were retained. For comparison, we evaluated logistic regression as a linear baseline under the same LOEO protocol, feature subsets, and lookback windows. The remaining missing values were then median-imputed and features standardized using training-fold statistics, followed by training with scikit-learn's default L2 regularization ($C=1.0$) without additional tuning~\cite{pedregosa2011scikit}.

To obtain a compact and interpretable feature set, feature selection was performed once using a fixed development split (split 0) and the 730-day lookback. Within each LOEO setting, features were ranked on the non-held-out training folds by gain importance, with split importance used as a secondary criterion, and the rankings were aggregated across the four settings. Highly correlated feature pairs ($\vert{}\rho\vert{} \geq 0.9$, Spearman) were pruned by discarding the lower ranked feature, reducing the candidate pool from $58$ to $38$. By applying the one-standard-error rule to validation PR-AUC, together with a constraint requiring coverage of all eight groups, we identified a final set of 17 features (Table~\ref{tab:full_features_final}) without using any test partition. The coverage constraint was intended to preserve representation across the proposed behavioral taxonomy rather than to maximize predictive performance alone. To assess performance across feature-set sizes, we constructed nested subsets with \(k \in \{4,8,12,17\}\), following the aggregated ranking while progressively expanding taxonomy coverage. Combined with the 100 base LOEO split--lookback configurations, these subsets produced \(4 \times 5 \times 5 \times 4 = 400\) model settings.

Hyperparameter optimization (HPO) was conducted once using Optuna over 30 trials~\cite{akiba2019optuna}, under the fixed development setting, the 730-day lookback, and \(k=17\). For each trial, validation PR-AUC was calculated separately for the four LOEO tasks corresponding to Binance, Bitget, OKX, and Korbit, and their unweighted mean was used as the optimization objective. The resulting hyperparameters were fixed across all five splits, lookback windows, and feature-set sizes in the main experiments. This procedure avoids setting-specific retuning and enables consistent model comparisons across feature-set sizes and temporal lookbacks. The selected configuration used \texttt{num\_leaves}=106, \texttt{learning\_rate}=0.0539, \texttt{subsample}=0.952,  \texttt{min\_child\_samples}=243, \texttt{colsample\_bytree}=0.735, and \(\lambda_{L2}=0.356\). We used 1,000 boosting rounds as an upper bound, with the effective number of rounds determined by validation-based early stopping with a patience of 50 rounds.


%% file: 6_discussion.tex
\section{Results}

\subsection{Leave-One-Exchange-Out Results}






\begin{table*}[t!]
\centering
\caption{Main LOEO performance comparison between the logistic regression (LR) baseline and LightGBM across feature-set sizes ($k$) and lookback windows ($L$), measured by macro PR-AUC with 95\% CIs.}
\label{tab:main_loeo_pr_auc}

{\renewcommand{\arraystretch}{1.1}
\setlength{\tabcolsep}{6.5pt}
\begin{tabular}{ccccccc}
\toprule
\multirow{2}{*}{\textbf{Features ($k$)}}
& \multirow{2}{*}{\textbf{Method}}
& \multicolumn{5}{c}{\textbf{Lookback window ($L$)}} \\
\cmidrule(lr){3-7}
& & \textbf{30 (1M)}
& \textbf{90 (3M)}
& \textbf{180 (6M)}
& \textbf{365 (1Y)}
& \textbf{730 (2Y)} \\
\midrule

\multirow{2}{*}{4}
& LR
& 0.3543 $\pm$ 0.0049
& 0.3809 $\pm$ 0.0069
& 0.4207 $\pm$ 0.0020
& 0.4862 $\pm$ 0.0020
& 0.6692 $\pm$ 0.0009 \\
& LightGBM
& 0.6003 $\pm$ 0.0033
& 0.6491 $\pm$ 0.0042
& 0.6945 $\pm$ 0.0065
& 0.7832 $\pm$ 0.0045
& 0.9052 $\pm$ 0.0030 \\
\cmidrule[0.05pt](lr){1-7}

\multirow{2}{*}{8}
& LR
& 0.3266 $\pm$ 0.0065
& 0.3580 $\pm$ 0.0043
& 0.4049 $\pm$ 0.0008
& 0.4731 $\pm$ 0.0010
& 0.6582 $\pm$ 0.0012 \\
& LightGBM
& 0.6253 $\pm$ 0.0019
& 0.6856 $\pm$ 0.0018
& 0.7368 $\pm$ 0.0029
& 0.8178 $\pm$ 0.0023
& 0.9212 $\pm$ 0.0045 \\
\cmidrule[0.05pt](lr){1-7}

\multirow{2}{*}{12}
& LR
& 0.4671 $\pm$ 0.0054
& 0.5216 $\pm$ 0.0133
& 0.5608 $\pm$ 0.0048
& 0.5971 $\pm$ 0.0013
& 0.7003 $\pm$ 0.0055 \\
& LightGBM
& 0.6518 $\pm$ 0.0033
& 0.7256 $\pm$ 0.0007
& 0.7829 $\pm$ 0.0024
& 0.8700 $\pm$ 0.0043
& 0.9487 $\pm$ 0.0028 \\
\cmidrule[0.05pt](lr){1-7}

\multirow{2}{*}{17}
& LR
& 0.5322 $\pm$ 0.0011
& 0.5656 $\pm$ 0.0009
& 0.5955 $\pm$ 0.0012
& 0.6553 $\pm$ 0.0012
& 0.7697 $\pm$ 0.0012 \\
& LightGBM
& 0.6611 $\pm$ 0.0052
& 0.7457 $\pm$ 0.0031
& 0.8049 $\pm$ 0.0044
& 0.8875 $\pm$ 0.0037
& \textbf{0.9594 $\pm$ 0.0017} \\
\bottomrule
\end{tabular}}
\vspace{1.5mm}
\end{table*}

\begin{figure}[b!]
    \centering
    \includegraphics[width=1.00\linewidth]{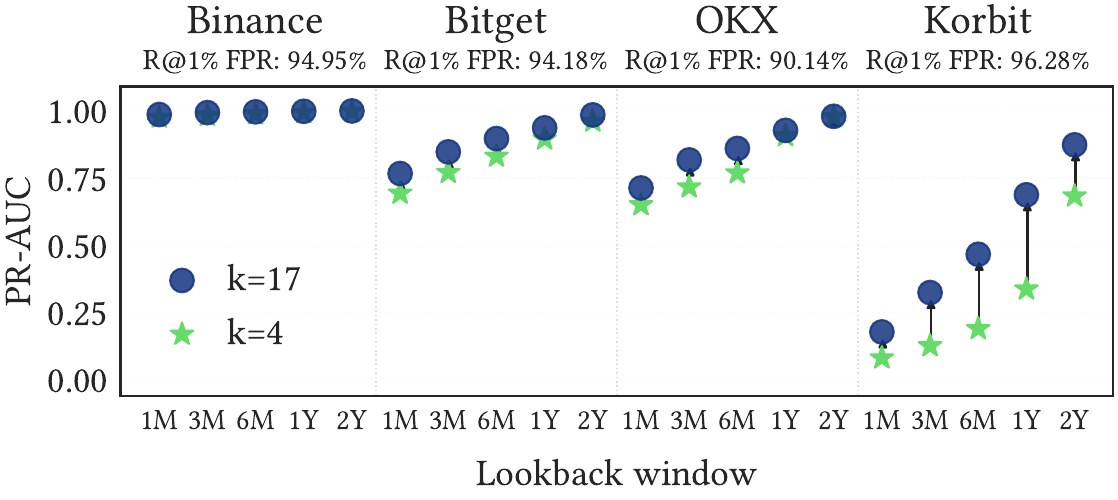}
    \caption{Per-exchange LOEO PR-AUC across lookback windows for the $k=4$ and $k=17$ feature sets. R@1\% FPR denotes recall at a validation-calibrated 1\% FPR ($k=17$, $L=730$).}
    \label{fig:per_exchange_performance}
\end{figure}

To evaluate cross-exchange generalization, we analyzed macro PR-AUC across all combinations of feature-set sizes ($k \in \{4, 8, 12, 17\}$) and lookback windows ($L \in \{30, 90, 180, 365, 730\}$ days, ranging from one month to two years). For each configuration, macro PR-AUC was computed as the unweighted mean across the four held-out exchanges within each split and then averaged over the five splits. The 95\% confidence intervals were calculated from the five split-level macro PR-AUC values. As summarized in Table~\ref{tab:main_loeo_pr_auc}, LightGBM consistently outperformed the LR baseline across all feature-set sizes and lookback windows. LightGBM's performance improved monotonically over the tested grid along both the feature and lookback dimensions, whereas LR showed a decline when increasing $k$ from 4 to 8 across all lookback windows. Under the most constrained setting ($k=4, L=30$), LightGBM achieved a macro PR-AUC of 0.6003. Extending the lookback window to $L=730$ at $k=4$ raised performance to 0.9052, whereas expanding the feature set to $k=17$ at $L=30$ yielded a more modest increase to 0.6611. The best performance was obtained using all 17 features and the 730-day lookback window, with a macro PR-AUC of 0.9594, compared with 0.7697 for LR under the same configuration. The small dispersion across repeated partitions suggests limited sensitivity to address-level partitioning. Overall, the results show that CEX identification benefits from broader feature coverage and longer lookback windows under the evaluated protocol, with the largest gains observed when both were increased.

Figure~\ref{fig:per_exchange_performance} decomposes performance by held-out exchange for the smallest and largest feature sets. PR-AUC remained high for Binance, while Bitget and OKX improved with longer lookback windows and the larger feature set. The largest gain was observed for Korbit, where PR-AUC increased from 0.0810 at $k=4$ and $L=30$ to 0.8735 at $k=17$ and $L=730$. Because the random PR-AUC baseline equals the proportion of CEX addresses in each test set, it differs across held-out exchanges: 0.8347 for Binance, 0.1729 for Bitget, 0.1423 for OKX, and 0.0083 for Korbit. Accordingly, per-exchange PR-AUC values should be interpreted relative to these baselines rather than compared solely by their absolute magnitudes. To complement these prevalence-sensitive results, we report recall using a threshold calibrated to a 1\% false-positive rate (FPR) on validation Non-CEX data. Under the final configuration, recall was high across all held-out exchanges (90.14–96.28\%), including Korbit, which showed the lowest PR-AUC. The held-out results suggest that the classifier can identify addresses from exchanges beyond Binance, despite its dominance in the CEX dataset.

\begin{table}[b!]
    \centering
    \caption{Selected features by mean absolute SHAP value across 20 LOEO models, evaluated on held-out test samples.}
    \label{tab:feature_interpretation}
    \resizebox{\columnwidth}{!}{%
    \begin{tabular}{r l c r r r c}
    \toprule
    \# & Feature & Group & $|\mathrm{SHAP}|$ & Share & Cum. & $\uparrow$ pushes \\
    \midrule
    1  & \texttt{avg\_gas\_used}              & G6 & 3.238 & 23.8\% & 23.8\%  & Non-CEX \\
    2  & \texttt{internal\_tx\_count\_out}    & G7 & 1.453 & 10.7\% & 34.4\%  & Non-CEX \\
    3  & \texttt{avg\_priority\_fee\_gwei}    & G6 & 1.451 & 10.6\% & 45.1\%  & CEX \\
    4  & \texttt{unique\_receivers}           & G4 & 1.269 &  9.3\% & 54.4\%  & Non-CEX \\
    5  & \texttt{avg\_subtraces}              & G7 & 0.997 &  7.3\% & 61.7\%  & Non-CEX \\
    6  & \texttt{eth\_net\_flow}              & G2 & 0.962 &  7.1\% & 68.7\%  & CEX \\
    7  & \texttt{erc20\_to\_eth\_tx\_ratio}   & G3 & 0.550 &  4.0\% & 72.8\%  & Non-CEX$^\dagger$ \\
    8  & \texttt{eth\_recv\_max}              & G1 & 0.504 &  3.7\% & 76.5\%  & CEX \\
    9  & \texttt{eth\_sent\_p95}              & G1 & 0.446 &  3.3\% & 79.7\%  & CEX$^\dagger$ \\
    10 & \texttt{eth\_sent\_median}           & G1 & 0.433 &  3.2\% & 82.9\%  & Non-CEX \\
    11 & \texttt{erc20\_unique\_tokens\_sent} & G3 & 0.427 &  3.1\% & 86.0\%  & CEX$^\dagger$ \\
    12 & \texttt{stablecoin\_tx\_ratio}       & G8 & 0.398 &  2.9\% & 89.0\%  & CEX \\
    13 & \texttt{forwarding\_value\_ratio}    & G8 & 0.389 &  2.9\% & 91.8\%  & Non-CEX \\
    14 & \texttt{avg\_fee\_eth}               & G6 & 0.388 &  2.8\% & 94.7\%  & CEX$^\dagger$ \\
    15 & \texttt{pct\_tx\_to\_contract}       & G5 & 0.319 &  2.3\% & 97.0\%  & CEX$^\dagger$ \\
    16 & \texttt{erc20\_tx\_count\_in}        & G3 & 0.264 &  1.9\% & 98.9\%  & CEX \\
    17 & \texttt{zero\_value\_tx\_ratio}      & G2 & 0.144 &  1.1\% & 100.0\% & Non-CEX \\
    \bottomrule
    \addlinespace

\multicolumn{7}{r}{
    \parbox{0.9\linewidth}{
        \raggedleft
        $^\dagger$ \normalsize
        Correlation sign differs across held-out exchanges.
    }
}
    \end{tabular}%
    }
  \end{table}

To interpret the best-performing configuration (\(k=17\), \(L=730\)), Table~\ref{tab:feature_interpretation} reports mean absolute SHAP values for the 17 selected features, computed on held-out test samples and averaged equally across 20 LOEO models. The two most influential features were \texttt{avg\_gas\_used} and \texttt{internal\_tx\_count\_out}, accounting for 23.8\% and 10.7\% of the total attribution, respectively, followed by \texttt{avg\_priority\_fee\_gwei} (10.6\%). Together, these features highlight the importance of gas behavior and internal transaction activity in model predictions. We further examined cross-exchange consistency in feature importance and directional associations. Among held-out CEX addresses, gas behavior (G6) had the highest aggregate SHAP importance across all four exchanges, although the leading individual feature differed: \texttt{avg\_gas\_used} ranked first for Binance, OKX, and Korbit, whereas \texttt{avg\_priority\_fee\_gwei} ranked first for Bitget. Directional associations were summarized using Spearman correlations between feature values and their CEX-class SHAP contributions; positive and negative signs indicate associations with CEX- and Non-CEX-directed contributions, respectively, as feature values increase. The correlation signs agreed across all four exchanges for 12 of the 17 selected features, including the six highest-ranked features, while the remaining five showed exchange-dependent differences and accounted for 15.5\% of total attribution. Collectively, these results suggest that the models relied on multiple behavioral dimensions, including gas and fee behavior, transaction structure, and asset-flow patterns, with some variation across exchanges. These attributions describe model predictions within the constructed dataset rather than intrinsic CEX-specific behavior.

\begin{table*}[t!]
\centering
\caption{Robustness and feature-group ablation checks using macro PR-AUC, with $\Delta$ relative to each baseline.}
\label{tab:robustness_ablation}

\small
\setlength{\tabcolsep}{3.0pt}
\renewcommand{\arraystretch}{0.94}

\begin{minipage}[t]{0.48\textwidth}
\centering
\begin{tabularx}{\linewidth}{@{}X r r r@{}}
\multicolumn{4}{c}{\textbf{Robustness checks}} \\[0.2em]
\toprule
Variant & $N$ & PR-AUC & $\Delta$ \\
\midrule
\addlinespace[0.45em]
Raw baseline & 3.03M & $0.9594 \pm 0.0017$ & 0.0000 \\
\quad Split- and exchange-nested selection & 3.03M & $0.9565 \pm 0.0037$ & -0.0029 \\
\quad Exchange-balanced reweighting & 3.03M & $0.9667 \pm 0.0009$ & +0.0073 \\
\quad No-additional-history excluded & 1.97M & $0.9683 \pm 0.0009$ & +0.0089 \\
\quad No-additional-history only & 1.06M & $0.9485 \pm 0.0025$ & -0.0109 \\
\quad Last active before 2022 & 0.30M & \(0.8174 \pm 0.0426\) & -0.1420 \\
\quad Last active from 2022 & 2.73M & \(0.9604 \pm 0.0016\) & +0.0010 \\
\quad Last active before 2025 & 1.80M & \(0.9509 \pm 0.0018\) & -0.0085 \\
\quad Last active from 2025 & 1.23M & \(0.9581 \pm 0.0010\) & -0.0013 \\

\addlinespace[0.05em]
\bottomrule
\end{tabularx}
\end{minipage}%
\hfill%
\begin{minipage}[t]{0.49\textwidth}
\centering
\begin{tabularx}{\linewidth}{@{}X r r r r@{}}
\multicolumn{5}{c}{\textbf{Feature-group ablation}} \\[0.2em]
\toprule
Removed group & $k'$ & PR-AUC & $\Delta$ & 95\% CI of $\Delta$ \\
\midrule
Full model & 17 & 0.9594 & 0.0000 & -- \\
\addlinespace[0.18em]
\quad G1. Transaction statistics & 14 & 0.9642 & +0.0048 & [+0.0044, +0.0053] \\
\quad G2. ETH flow patterns & 15 & 0.9617 & +0.0023 & [+0.0018, +0.0028] \\
\quad G3. ERC-20 token activity & 14 & 0.9468 & -0.0125 & [-0.0132, -0.0118] \\
\quad G4. Counterparty patterns & 16 & 0.9568 & -0.0026 & [-0.0030, -0.0022] \\
\quad G5. Contract interactions & 16 & 0.9588 & -0.0005 & [-0.0008, -0.0003] \\
\quad G6. Gas usage \& fees & 14 & 0.9305 & -0.0289 & [-0.0304, -0.0273] \\
\quad G7. Internal transactions & 15 & 0.9525 & -0.0069 & [-0.0077, -0.0061] \\
\quad G8. Composite patterns & 15 & 0.9575 & -0.0019 & [-0.0023, -0.0015] \\
\bottomrule
\end{tabularx}
\end{minipage}

\renewcommand{\arraystretch}{1.0}
\vspace{1.5mm}
\end{table*}

\subsection{Evaluation on Additional CEXs}

Extending our evaluation beyond the LOEO exchanges, we assessed two CEX-only datasets that did not overlap with the data used for model development or testing. \textit{Others} comprised 209 proof-of-reserves addresses from smaller exchanges omitted from LOEO because of limited sample sizes. \textit{External CEXs} contained 656 addresses manually collected from CEX-labeled entries in the top-holder lists of 45 tokens on Etherscan and Arkham, allowing each address to be linked to a specific exchange. Although these third-party labels are less direct than PoR disclosures, they broaden coverage and provide wallet role annotations for hot, cold, deposit, and unspecified addresses. Under \(k=17\) and \(L=730\), CEX scores were averaged across five split-specific models. Because both sets contained only CEX addresses, evaluation focused on score distributions and recall at a validation-defined 1\% false-positive rate threshold.

\begin{figure}[b!]
    \centering
    \includegraphics[width=1.00\linewidth]{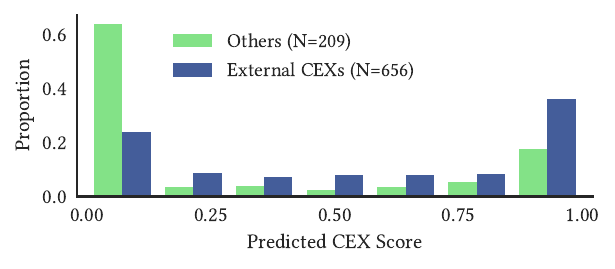}
    \vspace{-5.0mm}
    \caption{Predicted CEX score distributions for additional CEX-only evaluation sets under $k=17$ and $L=730$.}
    \label{fig:id_ood_score_single}
\end{figure}

As shown in Figure~\ref{fig:id_ood_score_single}, \textit{Others} exhibited lower scores (mean: $0.2719$, median: $0.0230$), yet achieved $59.90\%$ recall at the $1\%$ FPR threshold, reflecting partial feature overlap with exchange entities. In contrast, \textit{External CEXs} demonstrated relatively higher score alignment, with a mean of $0.5585$, a median of $0.6029$, and $90.34\%$ recall at the same threshold. Exchange-level analysis revealed high recall for Gate ($98.68\%$) and Coinbase ($84.31\%$), but lower recall for KuCoin ($56.25\%$) and Bitkub ($50.00\%$). Recall was also higher for hot wallets ($95.82\%$) and deposit addresses ($92.82\%$) than for cold ($83.70\%$) and unspecified addresses ($80.43\%$). These results indicate partial generalization to unseen CEX addresses, but the substantial variation across exchanges and wallet roles highlights that the model does not yet provide uniformly reliable coverage of heterogeneous CEX operations.

\subsection{Robustness Checks and Ablation}

Table~\ref{tab:robustness_ablation} (left) reports robustness checks for the final configuration. To assess potential bias from the global model selection procedure, we repeated feature selection and HPO independently for each of the 20 combinations of splits and held-out exchanges, using only the corresponding training and validation data rather than reusing the results from split 0. This yielded a macro PR-AUC of 0.9565 ($\Delta = -0.0029$), close to the baseline. We also examined the influence of Binance's large share of training addresses by reweighting CEX training samples to assign equal total weight to each exchange included in training, while preserving the overall CEX weight and leaving Non-CEX weights unchanged. Under the original validation and test procedures, macro PR-AUC increased to 0.9667, with the largest gain observed for Korbit ($\Delta = +0.0305$, 95\% CI [0.0280, 0.0330]). A complementary Binance-capped downsampling experiment showed a similar Korbit improvement ($\Delta = +0.0288$) relative to a random CEX control of equal size that approximately preserved the original exchange proportions, although the effects of balancing varied across exchanges.

Sensitivity to transaction history composition was examined by separating addresses with no additional transactions between the 30- and 730-day boundaries, defined by \(\text{tx\_count\_total}_{L=30}=\text{tx\_count\_total}_{L=730}\). PR-AUC remained high when these addresses were excluded (0.9683) or evaluated alone (0.9485), although these differences may reflect changes in subset composition rather than direct gains or losses. Using 2022 and 2025 as last activity cutoffs, performance remained broadly comparable to the baseline across three of the four subsets, with a notable decline only for addresses last active before 2022 (0.8174). This subset was smaller and highly skewed, with CEX addresses dominated by Binance and Non-CEX addresses by lending protocols. The lower performance may therefore reflect differences in both activity recency and address composition, making it difficult to determine how much each factor contributes to the observed decline.

To isolate the behavioral drivers of detection accuracy, we performed a LOEO ablation study on the full model ($k=17, L=730$). Feature groups were removed one at a time, and $\Delta$ was computed relative to the full model using paired bootstrap confidence intervals on the same test samples. As detailed in Table~\ref{tab:robustness_ablation} (right), removing gas behavior (G6) produced the largest performance drop ($\Delta = -0.0289$), followed by ERC-20 token activity (G3, $\Delta = -0.0125$) and internal transactions (G7, $\Delta = -0.0069$). Removing counterparty structure (G4), contract interaction (G5), and composite patterns (G8) also produced smaller negative changes. By contrast, excluding transaction statistics (G1) or ETH flow patterns (G2) slightly increased PR-AUC, suggesting that these groups did not provide additional predictive benefit under the evaluated configuration. These results highlight a potential trade-off between taxonomy coverage and predictive performance. However, these gains do not imply that the removed groups are uninformative, as related signals may be captured by features such as \texttt{forwarding\_value\_ratio} (G8). The paired confidence intervals excluded zero for all group removals, although several effect sizes were small. Taken together, the ablation results complement the robustness checks by showing that removing gas behavior, ERC-20 token activity, and internal transactions resulted in the largest decreases in predictive performance.


%% file: 7_conclusion.tex
\section{Conclusion and Limitations}

This research introduces a classification framework to identify centralized exchange addresses on Ethereum from on-chain behavioral features and verifiable proof-of-reserves labels. We construct an address-level dataset of 3,033,857 wallets spanning labeled CEX and heterogeneous Non-CEX entities, and define 58 domain-specific features covering transaction statistics, ETH and ERC-20 flows, counterparty structure, contract interaction, gas behavior, internal execution, and composite activity patterns. From these features, a validation-only selection procedure identified a compact set of 17 features used for the final models. Under a leave-one-exchange-out protocol across Binance, Bitget, OKX, and Korbit, macro PR-AUC increased from 0.6003 in the most constrained setting (\(k=4, L=30\)) to 0.9594 with the full selected feature set and a 730-day lookback window. The SHAP-based interpretation and ablation analyses showed that the model relied on multiple behavioral dimensions, with gas behavior, ERC-20 token activity, and internal transaction features providing the strongest contributions.

Evaluation on additional CEX-only datasets showed that this transfer was meaningful but not uniform. At a 1\% FPR threshold, the model recovered 90.34\% of externally collected CEX addresses, but recall was lower for the smaller PoR-based \textit{Others} group (59.90\%) and varied across exchanges and wallet roles. The robustness analyses showed broadly comparable performance across the evaluated alternative settings, providing additional support for the main findings. Overall, the results support an interpretable behavioral approach to CEX identification, aligning with the broader shift from heuristic clustering~\cite{meiklejohn2013fistful,victor2020address} toward machine learning-based behavioral signatures~\cite{garin2023machine}. Rather than relying on explicit transaction-link relationships, the framework uses aggregated address-level behavior, providing a complementary perspective when practices such as crypto dusting introduce ambiguous connections into transaction graphs~\cite{liu2026evasion, haffke2020cryptocurrencies}. The observed contributions of fee-related and transaction-flow features further highlight their predictive relevance to CEX address classification. From a practical perspective, the model should be viewed as a behavioral screening tool for AML and illicit-flow tracing studies~\cite{song2024identifying,wu2023toward}, where exchange attribution can help identify potential exit points and prioritize addresses for further investigation, rather than replace curated exchange labels.




Our findings point to several limitations and directions for future work. First, although the dataset combines high-confidence PoR-based CEX labels with five heterogeneous Non-CEX categories, its coverage remains limited by public disclosures and the negative sources considered. Including more non-exchange externally owned accounts (EOAs), custodial services, payment processors, and institutional wallets could better represent non-exchange activity. Exchange-balanced reweighting and downsampling checks maintained overall performance and improved results for some held-out exchanges, but the concentration of CEX addresses in Binance remains a limitation. SHAP and feature-group ablation identify predictive contributions within this dataset but cannot distinguish exchange-specific behavior from differences arising from the Non-CEX composition. Second, lookback windows are anchored to the first day of the month following each address's last observed transaction, so equal-length windows may span different calendar periods. Although the five windows allow us to examine the effect of historical coverage, the influence of calendar period differences cannot be fully isolated under this design. Future evaluations using fixed calendar periods and forward-in-time testing would help assess temporal robustness and generalization. Third, the analysis is restricted to Ethereum mainnet, leaving its applicability to other blockchain networks unexplored. Despite evaluation on held-out and additional CEXs, performance varied across exchanges and wallet roles, indicating that heterogeneous exchange operations remain challenging for the current address-level model. Given the challenges of early address detection with limited transaction histories~\cite{cheng2023evolve}, evaluating recently created CEX addresses would help determine its usefulness for AML investigations. Incorporating operational roles and graph-based relationships among deposit, hot, and cold wallets, including known sweeping patterns~\cite{skavysh2024market}, could support more robust entity-level attribution.

%% file: references.bib
@inproceedings{meiklejohn2013fistful,
  title={A {F}istful of {B}itcoins: {C}haracterizing {P}ayments {A}mong {M}en with {N}o {N}ames},
  author={Meiklejohn, Sarah and Pomarole, Marjori and Jordan, Grant and Levchenko, Kirill and McCoy, Damon and Voelker, Geoffrey M and Savage, Stefan},
  booktitle={Proceedings of the 2013 Conference on Internet Measurement Conference},
  pages={127--140},
  year={2013}
}

@article{ke2017lightgbm,
  title={Light{GBM}: {A} {H}ighly {E}fficient {G}radient {B}oosting {D}ecision {T}ree},
  author={Ke, Guolin and Meng, Qi and Finley, Thomas and Wang, Taifeng and Chen, Wei and Ma, Weidong and Ye, Qiwei and Liu, Tie-Yan},
  journal={Advances in Neural Information Processing Systems},
  volume={30},
  year={2017}
}

@article{vidal2025centralized,
  title={Centralized exchanges \& proof-of-solvency: The guardians of trust},
  author={Vidal-Tom{\'a}s, David},
  journal={Journal of International Financial Markets, Institutions and Money},
  volume={103},
  pages={102183},
  year={2025},
}

@inproceedings{victor2020address,
  title={Address clustering heuristics for {E}thereum},
  author={Victor, Friedhelm},
  booktitle={International Conference on Financial Cryptography and Data Security},
  pages={617--633},
  year={2020},
}

@inproceedings{victor2019measuring,
  title={{M}easuring {E}thereum-based {ERC}20 {T}oken {N}etworks},
  author={Victor, Friedhelm and L{\"u}ders, Bianca Katharina},
  booktitle={International Conference on Financial Cryptography and Data Security},
  pages={113--129},
  year={2019},
}

@article{chen2020understanding,
  title={{U}nderstanding {E}thereum via {G}raph {A}nalysis},
  author={Chen, Ting and Li, Zihao and Zhu, Yuxiao and Chen, Jiachi and Luo, Xiapu and Lui, John Chi-Shing and Lin, Xiaodong and Zhang, Xiaosong},
  journal={ACM Transactions on Internet Technology},
  volume={20},
  number={2},
  pages={1--32},
  year={2020},
}

@article{de2022structural,
  title={{T}he {S}tructural {R}ole of {S}mart {C}ontracts and {E}xchanges in the {C}entralisation of {E}thereum-{B}ased {C}ryptoassets},
  author={De Collibus, Francesco Maria and Pi{\v{s}}korec, Matija and Partida, Alberto and Tessone, Claudio J},
  journal={Entropy},
  volume={24},
  number={8},
  pages={1048},
  year={2022},
}

@inproceedings{beres2021blockchain,
  title={{B}lockchain is {W}atching {Y}ou: {P}rofiling and {D}eanonymizing {E}thereum {U}sers},
  author={B{\'e}res, Ferenc and Seres, Istv{\'a}n A and Bencz{\'u}r, Andr{\'a}s A and Quintyne-Collins, Mikerah},
  booktitle={2021 IEEE International Conference on Decentralized Applications and Infrastructures},
  pages={69--78},
  year={2021},
}

@techreport{reuter2025decrypting,
  title={{D}ecrypting {C}rypto: {H}ow to {E}stimate {I}nternational {S}tablecoin {F}lows},
  author={Reuter, Marco},
  year={2025},
  institution={International Monetary Fund (IMF)}
}

@article{chemaya2025dataset,
  title={A dataset of {U}niswap daily transaction indices by network},
  author={Chemaya, Nir and Cong, Lin William and Jorgensen, Emma and Liu, Dingyue and Zhang, Luyao},
  journal={Scientific Data},
  volume={12},
  number={1},
  pages={93},
  year={2025},
}

@inproceedings{akiba2019optuna,
  title={{O}ptuna: {A} {N}ext-generation {H}yperparameter {O}ptimization {F}ramework},
  author={Akiba, Takuya and Sano, Shotaro and Yanase, Toshihiko and Ohta, Takeru and Koyama, Masanori},
  booktitle={Proceedings of the 25th ACM SIGKDD International Conference on Knowledge Discovery \& Data Mining},
  pages={2623--2631},
  year={2019}
}

@article{grinsztajn2022tree,
  title={Why do tree-based models still outperform deep learning on typical tabular data?},
  author={Grinsztajn, L{\'e}o and Oyallon, Edouard and Varoquaux, Ga{\"e}l},
  journal={Advances in Neural Information Processing Systems},
  volume={35},
  pages={507--520},
  year={2022}
}

@article{pedregosa2011scikit,
  title={{S}cikit-learn: {M}achine {L}earning in {P}ython},
  author={Pedregosa, Fabian and Varoquaux, Ga{\"e}l and Gramfort, Alexandre and Michel, Vincent and Thirion, Bertrand and Grisel, Olivier and Blondel, Mathieu and Prettenhofer, Peter and Weiss, Ron and Dubourg, Vincent and others},
  journal={The Journal of Machine Learning Research},
  volume={12},
  pages={2825--2830},
  year={2011}
}

@article{lundberg2017unified,
  title={{A} {U}nified {A}pproach to {I}nterpreting {M}odel {P}redictions},
  author={Lundberg, Scott M and Lee, Su-In},
  journal={Advances in Neural Information Processing Systems},
  volume={30},
  year={2017}
}

@inproceedings{chen2021phishing,
  title={{P}hishing {S}cam {D}etection on {E}thereum: {T}owards {F}inancial {S}ecurity for {B}lockchain {E}cosystem},
  author={Chen, Weili and Guo, Xiongfeng and Chen, Zhiguang and Zheng, Zibin and Lu, Yutong},
  booktitle={Proceedings of the Twenty-Ninth International Joint Conferences on Artificial Intelligence},
  pages={4506--4512},
  year={2021}
}

@inproceedings{vasek2015there,
  title={{T}here’s {N}o {F}ree {L}unch, {E}ven {U}sing {B}itcoin: {T}racking the {P}opularity and {P}rofits of {V}irtual {C}urrency {S}cams},
  author={Vasek, Marie and Moore, Tyler},
  booktitle={International Conference on Financial Cryptography and Data Security},
  pages={44--61},
  year={2015},
}

@inproceedings{chen2018detecting,
  title={{D}etecting {P}onzi {S}chemes on {E}thereum: {T}owards {H}ealthier {B}lockchain {T}echnology},
  author={Chen, Weili and Zheng, Zibin and Cui, Jiahui and Ngai, Edith and Zheng, Peilin and Zhou, Yuren},
  booktitle={Proceedings of the 2018 World Wide Web Conference},
  pages={1409--1418},
  year={2018}
}

@inproceedings{weber2019anti,
  title={{A}nti-{M}oney {L}aundering in {B}itcoin: {E}xperimenting with {G}raph {C}onvolutional {N}etworks for {F}inancial {F}orensics},
  author={Weber, Mark and Domeniconi, Giacomo and Chen, Jie and Weidele, Daniel Karl I and Bellei, Claudio and Robinson, Tom and Leiserson, Charles E},
  booktitle={KDD '19 Workshop on Anomaly Detection in Finance},
  year={2019}
}

@inproceedings{lorenz2020machine,
  title={Machine learning methods to detect money laundering in the {B}itcoin blockchain in the presence of label scarcity},
  author={Lorenz, Joana and Silva, Maria In{\^e}s and Apar{\'\i}cio, David and Ascens{\~a}o, Jo{\~a}o Tiago and Bizarro, Pedro},
  booktitle={Proceedings of the First ACM International Conference on AI in Finance (ICAIF)},
  pages={1--8},
  year={2020}
}

@inproceedings{song2024identifying,
  title={{I}dentifying {M}oney {L}aundering {S}ubgraphs on the {B}lockchain},
  author={Song, Kiwhan and Dhraief, Mohamed Ali and Xu, Muhua and Cai, Locke and Chen, Xuhao and Mithal, Arvind and Chen, Jie},
  booktitle={Proceedings of the 5th ACM International Conference on AI in Finance (ICAIF)},
  pages={195--203},
  year={2024}
}

@article{wu2023toward,
  title={{T}oward {U}nderstanding {A}sset {F}lows in {C}rypto {M}oney {L}aundering {T}hrough the {L}enses of {E}thereum {H}eists},
  author={Wu, Jiajing and Lin, Dan and Fu, Qishuang and Yang, Shuo and Chen, Ting and Zheng, Zibin and Song, Bowen},
  journal={IEEE Transactions on Information Forensics and Security},
  volume={19},
  pages={1994--2009},
  year={2023},
}

@inproceedings{ron2013quantitative,
  title={{Q}uantitative {A}nalysis of the {F}ull {B}itcoin {T}ransaction {G}raph},
  author={Ron, Dorit and Shamir, Adi},
  booktitle={International Conference on Financial Cryptography and Data Security},
  pages={6--24},
  year={2013}
}

@article{lin2020modeling,
  title={{M}odeling and {U}nderstanding {E}thereum {T}ransaction {R}ecords via {A} {C}omplex {N}etwork {A}pproach},
  author={Lin, Dan and Wu, Jiajing and Yuan, Qi and Zheng, Zibin},
  journal={IEEE Transactions on Circuits and Systems II: Express Briefs},
  volume={67},
  number={11},
  pages={2737--2741},
  year={2020},
}

@article{garin2023machine,
  title={Machine learning in classifying bitcoin addresses},
  author={Garin, Leonid and Gisin, Vladimir},
  journal={The Journal of Finance and Data Science},
  volume={9},
  pages={100109},
  year={2023},
}

@article{hoang2022loaded,
  title={Loaded for bear: {B}itcoin private wallets, exchange reserves and prices},
  author={Hoang, Lai T and Baur, Dirk G},
  journal={Journal of Banking \& Finance},
  volume={144},
  pages={106622},
  year={2022},
}

@article{makarov2020trading,
  title={Trading and arbitrage in cryptocurrency markets},
  author={Makarov, Igor and Schoar, Antoinette},
  journal={Journal of Financial Economics},
  volume={135},
  number={2},
  pages={293--319},
  year={2020},
}

@inproceedings{iftikhar2025systemic,
  title={{S}ystemic {R}isk {A}ssessment in {CEX}s and {DEX}s: {A}n {A}pplication of a {T}ail {D}ependence-{B}ased {MST} and {C}o{V}a{R} {A}pproach},
  author={Iftikhar, Erum and Beccalli, Elena},
  booktitle={International Conference on Blockchain and Trustworthy Systems},
  pages={88--102},
  year={2025},
}

@article{skavysh2024market,
  title={{M}arket structure of cryptoasset exchanges: {I}ntroduction, challenges and emerging trends},
  author={Skavysh, Vladimir and Sharples, Jacob and Priazhkina, Sofia and Hasham, Salman H},
  year={2024},
  journal={Bank of Canada}
}

@article{haffke2020cryptocurrencies,
  title={{C}ryptocurrencies and anti-money laundering: the shortcomings of the fifth {AML} {D}irective ({EU}) and how to address them},
  author={Haffke, Lars and Fromberger, Mathias and Zimmermann, Patrick},
  journal={Journal of Banking Regulation},
  volume={21},
  number={2},
  pages={125--138},
  year={2020},
}

@inproceedings{cheng2023evolve,
  title={{E}volve {P}ath {T}racer: {E}arly {D}etection of {M}alicious {A}ddresses in {C}ryptocurrency},
  author={Cheng, Ling and Zhu, Feida and Wang, Yong and Liang, Ruicheng and Liu, Huiwen},
  booktitle={Proceedings of the 29th ACM SIGKDD Conference on Knowledge Discovery and Data Mining},
  pages={3889--3900},
  year={2023}
}

@inproceedings{liu2026evasion,
  title={{E}vasion {U}nder {B}lockchain {S}anctions},
  author={Liu, Endong and Ryan, Mark and Zhou, Liyi and Berrang, Pascal},
  booktitle={Proceedings of the ACM Web Conference 2026},
  pages={3507--3518},
  year={2026}
}
